\documentclass{llncs}

\usepackage[german,english]{babel}
\usepackage{times,theorem}
\usepackage{latexsym,color,graphics}
\usepackage{url}
\usepackage{epsfig}
\usepackage{amssymb}
\usepackage{amsmath}
\usepackage{amsfonts}
\usepackage{diagrams11}
\diagramstyle[tight,centredisplay%
,dpi=600
,UglyObsolete
,heads=LaTeX%
]

\begingroup
\catcode`\~=11
\gdef\urltilde{\lower 0.6ex\hbox{~}}
\endgroup

\newcommand{\A}{\mathcal{A}} \newcommand{\B}{\mathcal{B}}
\newcommand{\C}{\mathcal{C}} \newcommand{\D}{\mathcal{D}}
 \newcommand{\F}{\mathcal{F}}
 
\newcommand{\I}{\mathcal{I}} 
 
\newcommand{\M}{\mathcal{M}} 
 \renewcommand{\P}{\mathcal{P}}
 
\renewcommand{\S}{\mathcal{S}} \newcommand{\T}{\mathcal{T}}
 \newcommand{\V}{\mathcal{V}}

\title{Saturation of the morphisms in the database category}
\author{Zoran Majki\'c}
\authorrunning{Zoran Majki\'c}
\institute{ISRST, Tallahassee, FL, USA\\
\email{majk.1234@yahoo.com}\\ http://zoranmajkic.webs.com/}
\authorrunning{Zoran Majki\'c}

\newtheorem{coro}{Corollary}

\newcount\pdfoutput
\begin{document}


\maketitle

\begin{abstract}
In this paper we present the problem of saturation of a given
morphism in the database category \textbf{DB}, which is the base
category for the functiorial semantics of the database schema
mapping systems used in Data Integration theory. This phenomena
appears in the case when we are using the Second-Order
tuple-generating dependencies (SOtgd) with existentially quantified
non-built-in functions, for
the database schema mappings.\\
We provide the algorithm of the saturation for a given morphism,
which represents a mapping between two relational databases, and
show that the   original morphism  in \textbf{DB} can be
equivalently substituted by its more powerful saturated version in
any commutative diagram in \textbf{DB}.
\end{abstract}


\section{Introduction}
Since the late 1960s, there has been considerable progress in
understanding the algebraic semantics of logic and type theory,
particularly because of the development of categorical analysis of
most of the structures of interest to logicians. Although there have
been other algebraic approaches to logic, none has been as far
reaching in its aims and in its results as the categorical approach
has been. From a fairly modest beginning, categorical logic has
matured very nicely in the past four decades.\\ Categorical logic is
a branch of category theory within mathematics, adjacent to
mathematical logic but more notable for its connections to
theoretical computer science \cite{Jaco99}. In broad terms,
categorical logic represents both syntax and semantics by a
category, and an interpretation by a functor. The categorical
framework provides a rich conceptual background for logical and
type-theoretic constructions. The subject has been recognizable in
these terms since around 1970.\\ The recent monograph \cite{Majk14},
relevant to this paper, presents a categorical logic (denotational
semantics) for database schema mapping \emph{based on views} in a
very general framework for
 database-integration/exchange and peer-to-peer.
 The base database category $~\textbf{DB}~$  (instead of traditional $\textbf{Set}$ category),
 with objects instance-databases and with
 morphisms (mappings which are not simple functions) between them, is used at an \emph{instance level} as
  a proper semantic domain for a database mappings based on a set of complex query computations \cite{Majk14}.\\
   The higher logical \emph{schema level} of mappings between databases, usually written in some
 high expressive logical language (ex.~\cite{Lenz02,FKMP03}, GLAV (LAV and GAV), tuple
 generating dependency)  can then be  translated functorially into this base
 "computation" category. \\
The formal logical framework for the schema mappings is defined,
based on the second-order tuple generating dependencies (SOtgds),
with existentially quantified functional symbols. Each tgd is a
material implication from the conjunctive formula (with relational
symbols of a source schema, preceded with negation as well) into a
particular relational symbol of the target schema. It was provided
in \cite{Majk14} a number of algorithms which transform these
logical formulae into the algebraic structure based on the theory of
R-operads. The schema database integrity constraints are transformed
in similar way so that both, the schema mappings and schema
integrity-constraints, are
formally represented by R-operads.\\
A database mapping system is represented as a graph where the nodes
are the database schemas and the arrows are the schema mappings or
the integrity-constraints for schemas. This representation  is used
to define the database mapping sketches (small categories), based on
the fact that each schema has an identity arrow (mapping) and that
the mapping-arrows satisfy the associative low for the composition of them.\\
Each Tarski's interpretation of a  logical formulae (SOtgds), used
to specify the database mappings, results in the instance-database
mappings composed of a set of particular functions between the
source instance-database and the target instance-database. Thus, an
interpretation of a database-mapping system may be formally
represented as a functor from the sketch category (schema database
graph) into a  category where an object is an instance-database
(i.e., a set of relational tables) and an arrow is a  set of mapping
functions. This paper is an extension of the denotational semantics for the
database mappings presented in \cite{Majk14}.\\
The plan of this paper is the following: In Section 2 we present the
categorial logic an its functorial semantics used for the
denotational semantics of the schema mappings between RDBs, based on
the $\textbf{DB}$ category  \cite{Majk14}. Then, in Section 3 we
provide the algorithm for the saturation of the morphisms in the
category $\textbf{DB}$ and we show that the saturated morphism is
equal to the standard, functorially derived from a schema mapping,
morphisms. Then we present two significant examples how we can use
the saturation of the morphisms for the definition of 1:N
relationships between RDB tables and for the parsing of RDBS into
the intensional RDBs (IRDBs).
\section{Functorial semantics for database mappings \label{sec:DBmap}}
A \emph{database schema} is a pair $\A = (S_A , \Sigma_A)$ where
$S_A$ is
  a countable set of relational symbols (predicates in FOL) $r\in \mathbb{R}$
  with finite arity
   $n = ar(r) \geq 1$ ($~ar:\mathbb{R} \rightarrow \mathcal{N}$). A
   domain $\D$ is a nonempty finite set of individual
   symbols.
A relation symbol $r \in \mathbb{R}$ represents the \emph{relational
name} and can be used as an atom $r(\textbf{x})$ of FOL with
variables in $\textbf{x} = \langle x_1,...,x_{ar(r)}\rangle$ (taken
from a given set of variables $x_i \in \V$) assigned to its columns,
so that $\Sigma_A$ denotes a set of sentences (FOL formulae without
free variables) called \emph{integrity constraints}.
\\An \emph{instance-database} of a nonempty schema  $\A$ is given by
$A = (\A,I_T) = \{R =\|r\| = I_T(r) ~|~r \in S_A \}$ where $I_T$ is
a Tarski's FOL interpretation which satisfies \emph{all} integrity
constraints in $\Sigma_A$ and maps a relational symbol $r \in S_A$
into an n-ary relation $R=\|r\|\in A$.  Thus, an instance-database
$A$ is a set of n-ary relations, managed by relational database
systems. We denote by $r_\emptyset$ a nullary relational symbol
corresponding logically to a propositional symbol of an tautology,
such that $\bot = \|r_\emptyset\| = \{<>\}$ where $<>$ denotes the
empty tuple. We assume that $r_\emptyset$ is part of any database
schema $\A$.\\
 If $A$ is an instance-database and $\phi$ is a sentence
then we write $A\models \phi~$ to mean that $A$ satisfies $\phi$. If
$\Sigma$ is a set of sentences then we write $A \models \Sigma$ to
mean that $A\models \phi$  for every sentence $\phi \in \Sigma$.
Thus the set of all instances of $\A$ is defined by $Inst(\A) = \{
A~|~ A \models \Sigma_A \}$.\\
We consider a rule-based \emph{conjunctive query} over a database
schema $\A$ as an expression $ q(\textbf{x})\longleftarrow
r_1(\textbf{u}_1), ..., r_n(\textbf{u}_n)$, with finite $n\geq 0$,
$r_i$ are the relational symbols (at least one) in $\A$ or the
built-in predicates (e.g. $\leq, =,$ etc.), $q$ is a relational
symbol not in $\A$ and $\textbf{u}_i$ are free tuples (i.e., one may
use either variables or constants). Recall that if $\textbf{v} =
(v_1,..,v_m)$ then $r(\textbf{v})$ is a shorthand for
$r(v_1,..,v_m)$. Finally, each variable occurring in $\textbf{x}$ is
a \emph{distinguished} variable that must also occur at least once
in $\textbf{u}_1,...,\textbf{u}_n$. Rule-based conjunctive queries
(called rules) are composed of a subexpression $r_1(\textbf{u}_1),
...., r_n(\textbf{u}_n)$ that is the \emph{body}, and
 the \emph{head} of this rule $q(\textbf{x})$. The deduced head-facts  of a conjunctive query $q(\textbf{x})$ defined over an instance $A$ (for a given Tarski's
 interpretation $I_T$ of  schema $\A$) are
 equal to $\|q(x_1,...,x_k)\| = \{<v_1,...,v_k> \in \D^k ~|~  A \models \exists \textbf{y}(r_1(\textbf{u}_1)\wedge
 ...\wedge
r_n(\textbf{u}_n))[x_i/v_i]_{1\leq i \leq k} \}
 = I_T^*(\exists \textbf{y}(r_1(\textbf{u}_1)\wedge ...\wedge
r_n(\textbf{u}_n)))$, where the $\textbf{y}$ is a set of variables
which are not in the head of query, and $I_T^*$ is the unique
extension of $I_T$ to all formulae.
  Each conjunctive query corresponds
to a "select-project-join" term $t(\textbf{x})$
 of SPRJU algebra obtained from the formula $\exists \textbf{y}(r_1(\textbf{u}_1)\wedge ...\wedge
r_n(\textbf{u}_n))$.\\
We consider a finitary \emph{view} as a union of a finite set $S$ of
conjunctive  queries with the same
 head $q(\textbf{x})$ over a schema $\A$, and from the equivalent
algebraic point of view, it is a "select-project-join-rename +
union" (SPJRU) finite-length term $t(\textbf{x})$ which corresponds
to union of the terms of conjunctive queries in $S$. In what follows
we will use the same notation for a FOL formula $q(\textbf{x})$ and
its equivalent algebraic SPJRU expression $t(\textbf{x})$. A
materialized view of an instance-database $A$ is an n-ary relation
$R = \bigcup_{q(\textbf{x}) \in S}\|q(\textbf{x})\|_A$.  We denote
the set of all  finitary materialized
views that can be obtained from an instance $A$ by $TA$.\\
 We consider that a \emph{mapping} between two database schemas $\A
= (S_A , \Sigma_A)$ and $\B = (S_B , \Sigma_B)$ is expressed by an
union of "conjunctive queries with the same head". \\Such mappings
are called "view-based mappings", defined by a set of FOL sentences
$~ \{\forall \textbf{x}_i (q_{Ai}(\textbf{x}_i) \Rightarrow
q_{Bi}(\textbf{y}_i)) | ~$with$~ \textbf{y}_i \subseteq
\textbf{x}_i, 1 \leq i \leq n \}$, where $\Rightarrow$ is the
logical implication between these conjunctive queries
$q_{Ai}(\textbf{x}_i)$ and $q_{Bi}(\textbf{x}_i)$, over the
databases $\A$ and $\B$, respectively. Schema mappings are often
specified by the source-to-target tuple-generating dependencies
(tgds), used to formalize a data exchange \cite{FKMP03}, and in the
data integration scenarios under a name "GLAV assertions"
\index{global-and-local-as-view (GLAV)} \cite{Lenz02,CCGL02}. A tgd
\index{tuple generating dependency (tgd)} is a logical sentence (FOL
\index{First-Order Logic (FOL)} formula without free variables)
which says that if some tuples satisfying certain equalities exist
in the relation, then some other tuples (possibly with some unknown
values) must also exist in
another specified relation.\\
An equality-generating dependency (egd) \index{equality generating
dependency (egd)} is a logical sentence which says that if some
tuples satisfying certain equalities exist in the relation, then
some values in these tuples must be equal. Functional dependencies
are egds of a special form, for example primary-key integrity
constraints. \index{integrity constraints} Thus, egds are only used
for the specification of integrity constraints of a single database
schema, which define
the set of possible models of this database. They are not used for inter-schema database mappings.\\
 These two classes of dependencies
together comprise the \emph{embedded implication dependencies} (EID)
\index{embedded implication dependencies (EID)} ~\cite{Fagi82} which
seem to include essentially all of the naturally-occurring
constraints on relational databases (we recall that the bold symbols
$\textbf{x}, \textbf{y},..$ denote a nonempty list of variables):
\begin{definition} \label{def:EID}
We introduce the following two kinds of EIDs ~\cite{Fagi82}:
\begin{enumerate}
  \item  A \emph{tuple-generating dependency (tgd)} \index{tuple generating dependency
(tgd)}
     $~~~~~~ \forall {\bf x}  (q_A(\textbf{x})~ \Rightarrow ~
      q_B(\textbf{x})),$\\
      where $q_A(\textbf{x})$ is an existentially quantified formula
      $\exists {\bf y}~\phi_A(\bf x,\bf y)$ and $q_B(\textbf{x})$ is
      an existentially quantified formula $\exists {\bf z} ~ \psi_A(\bf x,\bf
      z)$, and
      where the formulae $\phi_A(\textbf{x}, \textbf{y})$  and $\psi_A(\textbf{x},\textbf{z})$ are conjunctions of
     atomic formulae (conjunctive queries) over the given database schemas. We assume the safety condition, that is, that every \emph{distinguished} variable in $\textbf{x}$  appears in $q_A$.\\
     We will consider also the class of \emph{weakly-full} tgds for which
     query answering is decidable, i.e., when $q_B(\textbf{x})$ has no existentially
      quantified variables, and if each $y_i \in \bf y$ appears at most once in $\phi_A(\bf x,\bf y)$.
     \item  An \emph{equality-generating dependency} (egd)  \index{equality generating dependency (egd)}
     $ ~~~~~~\forall {\bf x} ~(q_A(\bf x)~ \Rightarrow ~( y \doteq z ))$,\\
     where  $q_A(\textbf{x})$ is a conjunction of
     atomic formulae over a given database schema, and $\textbf{y} =<y_1,..,y_k>, ~\textbf{z} = <z_1,..,z_k>~$ are among the
     variables in \textbf{x}, and $\textbf{y} \doteq \textbf{z}$ is a shorthand for
     the formula $(y_1 \doteq z_1) \wedge ...\wedge (y_k \doteq
     z_k)$ with the built-in binary identity predicate $\doteq$ of the FOL.
\end{enumerate}
\end{definition}
Note that a tgd $\forall \textbf{x}(\exists {\bf y}~\phi_A(\bf x,\bf
y) \Rightarrow \exists {\bf z} ~ \psi_A(\bf x,\bf z))$ is logically
equivalent to the formula $\forall \textbf{x}\forall
\textbf{y}(\phi_A(\bf x,\bf y) \Rightarrow \exists {\bf z} ~
\psi_A(\bf x,\bf z))$, i.e., to $\forall \textbf{x}_1(\phi_A(
\textbf{x}_1) \Rightarrow \exists {\bf z} ~ \psi_A(\bf x,\bf z))$
with the set of distinguished variables $\textbf{x} \subseteq
\textbf{x}_1$.\\
 We  use for the integrity constraints $\Sigma_A$ of a
database schema $\A$ both tgds and egds, while for the inter-schema
mappings, between a schema $\A = (S_A, \Sigma_A)$ and a schema $\B =
(S_B, \Sigma_B)$, only the tgds $~ \forall {\bf x} (q_A(\textbf{x})~
\Rightarrow ~ q_B(\textbf{x}))$. So called second-order tgds (SO
tgds), has been introduced in \cite{FKPT05} as follows:
\begin{definition} \cite{FKPT05} \label{def:SOtgd} \index{Second
Order tgd (SOtgd)} Let $\A$ be a source schema and $\B$ a target
schema. A second-order
tuple-generating dependency (SO tgd) is a formula of the form:\\
$\exists \textbf{f}((\forall \textbf{x}_1(\phi_1  \Rightarrow
\psi_1)) \wedge...\wedge (\forall \textbf{x}_n(\phi_n  \Rightarrow
\psi_n)))$, where
\begin{enumerate}
  \item Each member of the tuple $\textbf{f}$ is a functional
  symbol.
  \item Each $\phi_i$ is a conjunction of:\\
  - atomic formulae of the form $r_A(y_1,...,y_k)$, where $r_A \in S_A$ is a
  k-ary relational symbol of schema $\A$ and $y_1,...,y_k$ are variables in
  $\textbf{x}_i$, not necessarily distinct;\\
  - the formulae with conjunction and negation connectives and  with  built-in
  predicate's atoms
  of the form $t \odot t'$, $\odot \in \{\doteq, <,>,... \}$, where $t$ and $t'$ are the terms based on
  $\textbf{x}_i$, $\textbf{f}$ and constants.
  \item Each $\psi_i$ is a conjunction of atomic formulae
  $r_B(t_1,...,t_m)$ where $r_B \in S_B$ is an m-ary relational symbol of
  schema $\B$ and $t_1,...,t_m$ are terms based on
  $\textbf{x}_i$, $\textbf{f}$ and constants.
  \item Each variable in $\textbf{x}_i$ appears in some atomic
  formula of $\phi_i$.
\end{enumerate}
\end{definition}
Notice that each constant $\overline{a}$ in an atom on the left-hand
side of implications must be substituted by new fresh variable $y_i$
and by adding a conjunct  $(y_i = \overline{a})$ in the left-hand
side of this implication, so that such atoms will have only the
variables (condition 2 above). For the empty set of tgds, we will
use the SOtgd tautology $r_\emptyset \Rightarrow r_\emptyset$.
 The forth condition is a "safety"
condition, analogous to that made for (first-order) tgds. It is easy
to see that every tgd is equivalent to one SOtgd without equalities.
For example, let $\sigma$ be the tgd \index{tuple generating
dependency (tgd)} $\forall x_1...\forall x_m (\phi_A(x_1,...,x_m)
\Rightarrow \exists
y_1...\exists y_n \psi_B(x_1,....,x_m,y_1,...,y_n))$.\\
It is logically equivalent to the following SOtgd without
equalities, which is
obtained by Skolemizing existential quantifiers in $\sigma$:\\
$\exists f_1...\exists f_n (\forall x_1...\forall x_m
(\phi_A(x_1,...,x_m) \Rightarrow \psi_B(x_1,....,x_m,f_1(x_1,....,x_m),\\...,f_n(x_1,....,x_m))))$.\\
Given a finite set $S$ of tgds  of an inter-schema mapping, we can
find a \emph{single} SOtgd that is equivalent to $S$ by taking, for
each tgd $\sigma$ in $S$, a conjunct of the SOtgd to capture
$\sigma$ as described above (we use
disjoint sets of function symbols in each conjunct, as before).\\
The  simultaneous inductive definition of  the set $\T X$ of
\emph{terms}
is as follows:\\
   1. All variables  $X \subseteq \V$ and constants  are terms. \\
   2. If $~t_1,...,t_k$ are terms  and $f_i$ a k-ary functional symbol then $f_i(t_1,...,t_k)$ is a
   term.\\
 An assignment $g:\V \rightarrow \D$ for variables in $\V$ is
applied only to free variables in terms and formulae.  Such an
assignment $g \in \D^{\V}$ can be recursively uniquely extended into
the assignment $g^*:\T X \rightarrow \D$, where $\T X$ denotes the
set of all terms with variables in $X \subseteq \V$, by :\\
1. $g^*(t_k) = g(x) \in \D$ if the term $t_k$ is a variable $x \in
\V$.\\
2. $g^*(t_k) = c \in \D$ if the term $t_k$ is a constant (nullary
functional symbol)
$\overline{c}$, with $g^*(\overline{1}) = 1$ for the truth-constant $\overline{1}$.\\
3. $g^*(f_i(t_1,...,t_k)) = I_T(f_i)(g^*(t_1),...,g^*(t_k)) \in\D$,
where $I_T(f_i)$
is a function obtained by Tarski's interpretation of the functional symbol $f_i$.\\
We denote by $~t_k/g~$ (or $\phi/g$) the ground term (or formula)
without free variables, obtained by assignment $g$ from a term $t_k$
(or a formula $\phi$), and by  $\phi[x/t_k]$ the formula
obtained by  uniformly replacing $x$ by a term $t_k$ in $\phi$.\\
 In what follows we use the algorithm $MakeOperads$  in \cite{Majk14}
 in order to transform logical
schema mappings $\M_{AB} = \{\Phi\}:\A \rightarrow \B$ given by the
SOtgds $\Phi$ in Definition  \ref{def:SOtgd} into the algebraic
operads $\textbf{M}_{AB} = MakeOperad(\M_{AB}) = \{v_1\cdot
q_{A,1},...,v_n\cdot q_{A,1},1_{r_\emptyset}\}:\A \rightarrow \B$.
The basic idea of the operad's operations $v_i \in O(r',r_B)$ and
$q_{A,i} \in O(r_1,...,r_k, r')$, where $r_i, 1\leq i \leq k$ are
relational symbols of the source schema $\A = (S_A, \Sigma_A)$ and
$r_B$ is a relational symbol of the target schema $\B$, and $r'$ has
the same type as $r_B$, is to formalize algebraically a mapping from
the set of source relations $r_i$ into a target relation $r_B$.\\\\
\textbf{Example 1}: Schema $\A = (S_A, \emptyset)$  consists of a
unary relation \verb"EmpAcme" that represents the employees of Acme,
a unary relation \verb"EmpAjax" that represents the employees of
Ajax, and unary relation \verb"Local" that represents employees that
work in the local office of their company. Schema $\B = (S_B,
\emptyset)$ consists of a unary relation \verb"Emp" that represents
all employees, a unary relation \verb"Local1" that is intended to be
a copy of \verb"Local", and unary relation \verb"Over65" that is
intended to represent people over age 65. Schema $\C = (S_C,
\emptyset)$ consists of a binary relation \verb"Office" that
associates employees with office numbers and unary relation
\verb"CanRetire" that represents employees eligible for retirement.
Consider now the following schema mappings:\\
$\M_{AB} = \{\forall x_e(\verb"EmpAcme"(x_e)\Rightarrow
\verb"Emp"(x_e)) \wedge \forall x_e(\verb"EmpAjax"(x_e) \Rightarrow
\verb"Emp"(x_e)) \wedge \forall x_p(\verb"Local"(x_p)\Rightarrow
\verb"Local1"(x_p)) \}$, and \\ $\M_{BC} = \{\exists f_1(\forall
x_e((\verb"Emp"(x_e) \wedge \verb"Local1"(x_e)) \Rightarrow
\verb"Office"(x_e,f_1(x_e)) ) \wedge
\\ \forall x_e((\verb"Emp"(x_e) \wedge \verb"Over65"(x_e))
\Rightarrow \verb"CanRetire"(x_e))) \}$.\\
Then, by their composition, we obtain the
composed mapping $\M_{AC}:\A \rightarrow \C$ equal to\\
$\M_{AC} =  \{\exists f_1\exists f_2\exists f_{Over65}(\\\forall
x_e((\verb"EmpAcme"(x_e)\wedge \verb"Local"(x_e)) \Rightarrow
\verb"Office"(x_e,f_1(x_e)) \wedge \\ \forall
x_e((\verb"EmpAjax"(x_e)\wedge \verb"Local"(x_e)) \Rightarrow
\verb"Office"(x_e,f_2(x_e))) \wedge\\
 \forall x_e((\verb"EmpAcme"(x_e) \wedge (f_{Over65}(x_e) \doteq \overline{1}))
\Rightarrow \verb"CanRetire"(x_e)) \wedge \\ \forall
x_e((\verb"EmpAjax"(x_e)\wedge (f_{Over65}(x_e) \doteq
\overline{1})) \Rightarrow \verb"CanRetire"(x_e))
)\}$,\\
where $f_{Over65}$ is the characteristic function of the relation
(predicate) \verb"Over65" Which is not part of schema $\A$. Then, by
transformation into abstract operad's operations, we obtain
$\textbf{M}_{AC} = MakeOperads(\M_{AC}) = \{q^A_1, q^A_2, q^A_3,
q^A_4, 1_{r_\emptyset} \}$, $q^A_i  = v_i\cdot q_{A,i}$, where:\\
1. The operations $q^A_1 \in O(\verb"EmpAcme",\verb"Local",
\verb"Office")$ and $q_{A,1} \in O(\verb"EmpAcme",\verb"Local",
r'_1)$   correspond to  the expression $((\_~)_1(x_e)\wedge
(\_~)_2(x_e)) \Rightarrow (\_~)(x_e,f_1(x_e))$ and $v_1 \in O(r'_1,
\verb"Office")$ to $(\_~)_1(x_e,x_p) \Rightarrow
(\_~)(x_e,x_p)$;\\
2. The operations $q^A_2 \in O(\verb"EmpAjax",\verb"Local",
\verb"Office")$ and $q_{A,2} \in O(\verb"EmpAjax",\verb"Local",
r'_2)$ correspond
   to the expression $((\_~)_1(x_e)\wedge (\_~)_2(x_e)) \Rightarrow
(\_~)(x_e,f_2(x_e))$ and $v_2 \in O(r'_2, \verb"Office")$ to
$(\_~)_1(x_e,x_p) \Rightarrow
(\_~)(x_e,x_p)$;\\
3. The operations $q^A_3 \in O(\verb"EmpAcme",\verb"Over65",
\verb"CanRetire")$  and $q_{A,3} \in O(\verb"EmpAcme",\\
\verb"Over65", r'_3)$ correspond to the expression
$((\_~)_1(x_e)\wedge (\_~)_2(x_e)) \Rightarrow (\_~)(x_e)$  and $v_3
\in O(r'_3, \verb"CanRetire")$ to $(\_~)_1(x_e) \Rightarrow
(\_~)(x_e)$;\\
4. The operations $q^A_4 \in O(\verb"EmpAjax", \verb"Over65",
\verb"CanRetire")$ and $q_{A,4} \in O(\verb"EmpAjax",\\
\verb"Over65", r'_4)$ correspond to the expression
$((\_~)_1(x_e)\wedge (\_~)_2(x_e)) \Rightarrow (\_~)(x_e)$ and $v_4
\in O(r'_4, \verb"CanRetire")$ to $(\_~)_1(x_e) \Rightarrow
(\_~)(x_e)$.\\
These three arrows $\textbf{M}_{AB}:\A \rightarrow \B$,
$\textbf{M}_{BC}:\B \rightarrow \C$ and $\textbf{M}_{AC}:\A
\rightarrow \C$ compose a graph $G$ of this database mapping system.
 From
the fact that the operads can be composed, the composition of two
schema mappings $\textbf{M}_{AB}$ and $\textbf{M}_{BC}$ can be
translated into composition of operads which is associative, so that
they can be represented by the sketch category $\textbf{Sch}(G)$
derived from the graph $G$ of the schema mappings.
\\$\square$\\
Sketches are called
 graph-based logic and provide very clear and intuitive
 specification of computational data and activities. For any small sketch
 $\textbf{E}$, the category of models $Mod(\textbf{E})$ is an accessible category by Lair's theorem and
 reflexive subcategory of $\textbf{Set}^{\textbf{E}}$ by Ehresmann-Kennison theorem.
A generalization to  base categories other than $\textbf{Set}$ was
proved by Freyd and Kelly (1972) \cite{FrKe72}. The generalization
to $\textbf{DB}$ category is exhaustively provided in
\cite{Majk14}, so that the functorial semantics of a database mapping system
expressed by a graph $G$ is defined by a functor (R-algebra) $\alpha^*:\textbf{Sch}(G) \rightarrow \textbf{DB}$. \\
The R-algebra $\alpha$ is derived from a given Tarski's
interpretation $I_T$ of the given database schema mapping graph $G$
and represented by a sketch category $\textbf{Sch}(G)$ (with arrows
$M_{AB}:\A \rightarrow \B$, as in Example 1). R-algebra $\alpha$ is
equal to $I_T$ for the relations of the data schemas, $\alpha(r_i) =
I_T(r_i)$ is a relational table of the instance database $A =
\alpha^*(\A) = \{\alpha(r_i) ~|~r_i \in S_A\}$ ($\alpha^*$ denotes
the extension of $\alpha$ to sets), and $\alpha(q_{A,i}):\alpha(r_1)
\times...\times \alpha(r_k) \rightarrow \alpha(r')$ is a surjective
function from the relations in the instance database $A$ into its
image (relation) $\alpha(r')$, with a function
$\alpha(v_i):\alpha(r') \rightarrow \alpha(r_B)$ into the relation
of the instance database $B = \alpha^*(\B)$. \\
 We have
that for any R-algebra $\alpha$, $\alpha(r_\emptyset) = \bot =
\{<>\}$ is the empty relation composed by only empty tuple $<> \in
D_{-1}$, and $1_{r_\emptyset}$ is the identity operads operation of
the empty relation $r_\emptyset$, so that $q_\bot =
\alpha(1_{r_\emptyset}) = id_\bot:\bot \rightarrow \bot$ is the
identity function.\\\\
\textbf{Example 2}: For the operads defined in Example 1,
 let a mapping-interpretation \index{mapping-interpretations}(an R-algebra) $\alpha$  be an extension of Tarski's interpretation $I_T$ \index{Tarski's interpretations} of the
 source schema $\A = (S_A, \Sigma_A)$ that satisfies all constraints in $\Sigma_A$ and defines its database instance $A =
 \alpha^*(S_A) = \{\alpha(r_i)~|~r_i \in S_A \}$ and, analogously, an interpretation of
 $\C$.\\
 Let $\alpha$ satisfy the SOtgd of the mapping $\M_{AC}$ by the Tarski's interpretation for the
 functional symbols $f_i$,  for $1 \leq i \leq 2$, in this SOtgd (denoted by
 $I_T(f_i))$.\\
 Then we obtain the relations
 $\alpha$(\verb"EmpAcme"),  $\alpha$(\verb"EmpAjax"), $\alpha$(\verb"Local"),
 $\alpha$(\verb"Office") and $\alpha$(\verb"CanRetire"). The
 interpretation of $f_{Over65}$  is the characteristic function of
 the relation $\alpha(\verb"Over65"$ in the instance $B =
 \alpha^*(S_B)$ of the database $\B = (S_B,\Sigma_B)$, so that $\overline{f}_{Over65}(a) = 1~$
  if $~<a> ~\in \alpha(\verb"Over65")$).\\
  Then this mapping interpretation $\alpha$ defines the following
  functions:
  \begin{enumerate}
   \item The function $\alpha(q_{A,1}):\alpha(\verb"EmpAcme")\times \alpha(\verb"Local")
   \rightarrow \alpha(r'_1)$, such that for any tuple $<a>
   \in \alpha(\verb"EmpAcme")$ and $<b> \in \alpha(\verb"Local")$,\\
$\alpha(q_{A,1})(<a>,<b>) = <a,I_T(f_1(a))>~$ if $~a = b$; $~<>$ otherwise.\\
And for any $<a,b> \in \alpha(r'_1)$, $~\alpha(v_1)(<a,b>) = <a,b>$
if $ <a,b> \in \alpha(\verb"Office")$; $<>$ otherwise.
   \item The function $\alpha(q_{A,2}):\alpha(\verb"EmpAjax")\times \alpha(\verb"Local")
   \rightarrow \alpha(r'_2)$, such that for any tuple $<a>
   \in \alpha(\verb"EmpAjax")$ and $<b> \in \alpha(\verb"Local")$,\\
$\alpha(q_{A,2})(<a>,<b>) = <a,I_T(f_2(a))>~$ if $~a = b$; $~<>$ otherwise.\\
And for any $<a,b> \in \alpha(r'_2)$, $\alpha(v_2)(<a,b>) = <a,b>$
if $ <a,b> \in \alpha(\verb"Office")$; $<>$ otherwise.
   \item The function $\alpha(q_{A,3}):\alpha(\verb"EmpAcme")\times \alpha(\verb"Over65")\rightarrow \alpha(r'_3)$,
    such that for any tuple $<a>
   \in \alpha(\verb"EmpAcme")$ and $<b>
   \in \alpha(\verb"Over65")$,\\
$\alpha(q_{A,3})(<a>,<b>) = <a>~$, if $~a = b$;
 $~<>$ otherwise.\\
 And for any $<a> \in \alpha(r'_3)$, $\alpha(v_3)(<a>) =
<a>$ if $ <a> \in \alpha(\verb"CanRetire")$; $<>$ otherwise.
   \item The function $\alpha(q_{A,4}):\alpha(\verb"EmpAjax")\times \alpha(\verb"Over65") \rightarrow \alpha(r'_4)$,
    such that for any tuple $<a>
   \in \alpha(\verb"EmpAjax")$ and $<b>
   \in \alpha(\verb"Over65")$\\
$\alpha(q_{A,4})(<a>,<b>) = <a>~$, if $~a = b$;
 $~<>$ otherwise.\\
 And for any $<a> \in \alpha(r'_4)$, $\alpha(v_4)(<a>) =
<a>$ if $ <a> \in \alpha(\verb"CanRetire")$; $<>$ otherwise.
 \end{enumerate}
 From the fact that the mapping-interpretation satisfies the schema
 mappings, based on Corollary 4 in Section 2.4.1 \cite{Majk14}, all functions $\alpha(v_i)$, for $1
 \leq i \leq 4$, are the injections.
 \\$\square$\\
 Formal definition of an
R-algebra $\alpha$ as a mapping-interpretation of a schema mapping
$\M_{AB}:\A \rightarrow \B$  is given in \cite{Majk14} (Section
2.4.1, Definition 11) as follows:
\begin{definition} \label{def:mapping-interpretation}
Let $\phi_{Ai}(\textbf{x}) \Rightarrow r_B(\textbf{t})$ be an
implication $\chi$ in a normalized SOtgd  $\exists \textbf{f}(
\Psi)$ (where $\Psi$ is a FOL formula) of the mapping $\M_{AB}$,
$\textbf{t}$ be a tuple of terms with variables in $\textbf{x} =
<x_1,...,x_m>$, and  $q_i \in MakeOperads(\M_{AB})$ be the operad's
operation of this implication obtained by  $MakeOperads$ algorithm,
equal to the expression $(e \Rightarrow (\_~)(\textbf{t}))\in
O(r_1,...,r_k,r_B)$, where $q_i = v_i \cdot q_{A,i}$ with $q_{A,i}
\in O(r_1,...,r_k,r_q)$ and $v_i \in O(r_q, r_B)$ such that
 for a new relational symbol $r_q$,$~ar(r_q) = ar(r_B) \geq 1$.\\
Let $S$ be an empty set and $e[(\_~)_n/r_n]_{1\leq n \leq k}$ be the
formula obtained from expression $e$ where each place-symbol
$(\_~)_n$ is substituted by relational symbol $r_n$ for $1 \leq n
\leq k$. Then do the following as far as it is possible: For each
two relational symbols $r_j,r_n$ in the formula
$e[(\_~)_n/r_n]_{1\leq n \leq k}$ such that $j_h$-th \textsl{free
variable} (which is not an argument of a functional symbol) in the
atom $r_j(\textbf{t}_j)$ is equal to $n_h$-th free variable in the
atom $r_n(\textbf{t}_n)$ (both atoms in $e[(\_~)_n/r_n]_{1\leq n
\leq k})$, we insert the set $\{(j_h,j),(n_h,n)\}$ as one element of
$S$. At the end, $S$ is the set of sets that contain the pairs of mutually equal free variables.\\
 An R-algebra $\alpha$ is a \textsl{mapping-interpretation} of $\M_{AB}:\A \rightarrow \B$
 if it is
 an extension of  a Tarski's interpretation $I_T$, of all predicate and functional symbols in FOL formula
 $\Psi$, with $I_T^*$ being its extension  to all formulae), and if for each $q_i \in MakeOperads(\M_{AB})$ it satisfies the following:
\begin{enumerate}
  \item For each relational symbol
$r_i \neq r_\emptyset$ in $\A$ or $\B$,
   $\alpha(r_i) =   I_T(r_i)$.
  \item  We obtain a  function
$f = \alpha(q_{A,i}):R_1\times...\times R_k \rightarrow
\alpha(r_q)$,\\ where for each $1 \leq i \leq k$, $R_i =
\D^{ar(r_i)}\backslash \alpha(r_i)$
  if the place symbol $(\_~)_i \in q_i$ is preceded by negation
  operator $\neg$; $\alpha(r_i)$ otherwise,
such that for every $\textbf{d}_i \in R_i$:\\
$f(<\textbf{d}_1,...,\textbf{d}_k>) = g^*(\textbf{t}) =
~<g^*(t_1),...,g^*(t_{ar(r_B)})>~$\\
 if $~\bigwedge
\{\pi_{j_h}(\textbf{d}_j) =
\pi_{n_h}(\textbf{d}_n)~|~\{(j_h,j),(n_h,n)\} \in S \}~$ is true and
 the assignment $g$ satisfies the formula
$e[(\_~)_n/r_n]_{1\leq n \leq k}$; $~~<>$ (empty tuple)
otherwise,\\
where the assignment $g:\{x_1,...,x_m \}\rightarrow \D~$ is defined
by  the tuple of values $<g(x_1),...,g(x_m)> ~=
Cmp(S,<\textbf{d}_1,...,\textbf{d}_k>)$, and its extension $g^*$ to
all
terms  such that for any term $f_i(t_1,...,t_n)$:\\
$g^*(f_i(t_1,...,t_n))  = I_T(f_i)(g^*(t_1),...,g^*(t_n))$ if $n
\geq 1$; $~I_T(f_i)$ otherwise.\\
The algorithm $~Cmp$ (compacting the list of tuples by eliminating
the duplicates defined in $S$) is defined as
follows:\\
 \textbf{Input}: a set $S$ of joined (equal) variables defined above, and a list of tuples
 $<\textbf{d}_1,...,\textbf{d}_k>$.\\
 Initialize $\textbf{d}$ to $\textbf{d}_1$. Repeat consecutively the following,
 for $j=2,...,k$:\\
 Let $\textbf{d}_j$ by a tuple of values $<v_1,...,v_{j_n}>$, then for $i = 1,...,j_n$ repeat consecutively the
 following:\\
 $\textbf{d} = ~\textbf{d}~\& v_i~$ if there does not exist and element $\{(j_h,j),(n_h,n)\}$
 in
 $S$ such that $j \leq n$; $~\textbf{d}$, otherwise.\\
 (The operation of concatenation $'\&'$ appends the value $v_i$ at the end of tuple $\textbf{d}$)\\
\textbf{Output}: The tuple $~Cmp(S,<\textbf{d}_1,...,\textbf{d}_k>)
= \textbf{d}$.
\item  $\alpha(r_q)$  is equal to the image of the
function $f$ in point 2 above.
\item The function $h = \alpha(v_i):\alpha(r_q) \rightarrow
\alpha(r_B)$ such that for each $\textbf{b} \in \alpha(r_q)$,\\
$h(\textbf{b}) = \textbf{b}$ if $\textbf{b} \in \alpha(r_B)$; empty
tuple $<>$ otherwise.
\end{enumerate}
\end{definition}
Note that the formulae $\phi_{Ai}(\textbf{x})$ and expression
$e[(\_~)_n/r_n]_{1\leq n \leq k}$ are logically equivalent, with the
only difference that the atoms with characteristic functions
$f_r(\textbf{t}) \doteq \overline{1}$ in the first formula are
substituted by the atoms $r(\textbf{t})$, based on the fact that the
assignment $g$ satisfies $r(\textbf{t})$ iff $g^*(f_r(\textbf{t})) =
\overline{f}_r(g^*(\textbf{t})) = 1$ (and for every assignment
$g(\overline{1}) = 1$), where
 $\overline{f}_r:\D^{ar(r)} \rightarrow
\{0,1\}$  is the characteristic function of relation $\alpha(r)$
such that for each tuple $\textbf{c}\in \D^{ar(r)}$,
$\overline{f}_r(\textbf{c}) = 1$ if $\textbf{c} \in \alpha(r)$; $0$
otherwise.\\\\
\textbf{ Example 3}: Let us show  how we construct the set $S$ and
the
compacting of tuples given by Definition \ref{def:mapping-interpretation} above:\\
Let us consider an operad $q_i\in MakeOperads(\M_{AB})$, obtained
from a normalized implication $\phi_{Ai}(\textbf{x}) \Rightarrow
r_B(\textbf{t})$ in $\M_{AB}$, $((y \doteq f_1(x,z)) \wedge
r_1(x,y,z) \wedge  r_2(v,x,w) \wedge (f_{r_3}(y,z,w',w) \doteq
\overline{1})) \Rightarrow r_B(x,z,w,f_2(v,z))$, so that $q_i$ is
equal to the expression $(e \Rightarrow (\_)(\textbf{t})) \in
O(r_1,r_2,r_3,r_B)$, where $\textbf{x} = <x,y,z,v,w,w'>$ (the
ordering of variables in the atoms (with database relational
symbols) from left to right), $\textbf{t} =<x,z,w,f_2(v,z)> $, and
the expression $e$ equal to $ (y \doteq f_1(x,z)) \wedge
(\_~)_1(\textbf{t}_1) \wedge (\_~)_2(\textbf{t}_2) \wedge
(\_~)_3(\textbf{t}_3)$, with $\textbf{t}_1 = <x,y,z>$, $\textbf{t}_2
= <v,x,w>$ and
 $\textbf{t}_3 = <y,z,w',w>$.\\
Consequently, we obtain,\\ $S = \{\{(1,1),(2,2)\}, \{(2,1),(1,3)\},
\{(3,1),(2,3)\}, \{(3,2),(4,3)\} \}$,\\
that are the positions of duplicates (or joined variables) of
$x,y,z$ and $w$
respectively.\\
Thus, for  given tuples $\textbf{d}_1 = <a_1,a_2,a_3> \in
\alpha(r_1)$, $\textbf{d}_2 = <b_1,b_2,b_3> \in \alpha(r_2)$ and
$\textbf{d}_3 = <c_1,c_2,c_3,c_4> \in \alpha(r_3)$,  the statement
$~\bigwedge \{\pi_{j_h}(\textbf{d}_j) =
\pi_{n_h}(\textbf{d}_n)~|~\{(j_h,j),(n_h,n)\} \in S \}$ is equal to
$(\pi_1(\textbf{d}_1) = \pi_2(\textbf{d}_2)) \wedge
(\pi_2(\textbf{d}_1) = \pi_1(\textbf{d}_3)) \wedge
(\pi_3(\textbf{d}_1) = \pi_2(\textbf{d}_3)) \wedge
(\pi_3(\textbf{d}_2) = \pi_4(\textbf{d}_3))$, which is true when
$a_1 = b_2$, $a_2 = c_1$, $a_3 = c_2$ and $b_3 = c_4$.\\
The compacting of these tuples is equal to \\$\textbf{d} =
Cmp(S,<\textbf{d}_1,\textbf{d}_2,\textbf{d}_3>) =
<a_1,a_2,a_3,b_1,b_3,c_3>$, with the assignment to variables
$[x/a_1],
[y/a_2], [z/a_3], [v,b_1], [w/b_3]$ and $[w'/c_3]$.\\
That is,  $\textbf{d} =
\textbf{x}[x/a_1,y/a_2,z/a_3,v/b_1,w/b_3,w'/c_3]$ is obtained by
this assignment $g$ to the tuple of variables $\textbf{x}$, so that
the sentence $e[(\_~)_n/r_n]_{1\leq n \leq k}/g$ is
well defined and equal to:\\
$(a_2 = I_T(f_1)(a_1,a_3)) \wedge r_1(a_1,a_2,a_3) \wedge
r_2(b_1,a_1,b_3) \wedge r_3(a_2,a_3,c_3,b_3)$, that is  to\\
$(a_2 = I_T(f_1)(a_1,a_3)) \wedge r_1(\textbf{d}_1) \wedge
r_2(\textbf{d}_2) \wedge r_3(\textbf{d}_3)$, and if this formula is
satisfied by such an assignment $g$,  i.e.,
$I_T^*(e[(\_~)_n/r_n]_{1\leq n \leq k}/g) = 1$, then
\\$f(<\textbf{d}_1,\textbf{d}_2,\textbf{d}_3>) = g^*(\textbf{t})
= <g(x),g(z),g(w),g^*(f_2(v,z))> \\
= <a_1,a_3,b_3,I_T(f_2)(b_1,a_3)>$,
\\for a given Tarski's
interpretation $I_T$, where $I_T^*$ is the extension of $I_T$ to
all FOL formulae.\\
If $\M_{AB}$ is satisfied by the mapping-interpretation
$\alpha$,\index{mapping-interpretations} this value of
$f(<\textbf{d}_1,\textbf{d}_2,\textbf{d}_3>)$ corresponds to the
truth of the normalized implication in the SOtgd of $\M_{AB}$,
$\phi_{Ai}(\textbf{x}) \Rightarrow r_B(\textbf{t})$ for the
assignment $g$ derived by substitution $[\textbf{x}/\textbf{d}]$,
when $\phi_{Ai}(\textbf{x})/g$ is true. Hence, $r_B(\textbf{t})/g$
is equal to $r_B(<a_1,a_3,b_3,I_T(f_2)(b_1,a_3)>)$, i.e., to $
r_B(f(<\textbf{d}_1,\textbf{d}_2,\textbf{d}_3>))$ and has to be true
as well (i.e.
$I_T^*(r_B(f(<\textbf{d}_1,\textbf{d}_2,\textbf{d}_3>))) = 1$ or,
 equivalently, $f(<\textbf{d}_1,\textbf{d}_2,\textbf{d}_3>) \in \alpha(r_B) = I_T(r_B)$). \\
Consequently, if $\M_{AB}$ is satisfied by a mapping-interpretation
\index{mapping-interpretations} $\alpha$ (and hence $\alpha(v_i)$ is
an injection function with $\alpha(r_q) \subseteq \alpha(r_B)$) then
$f(<\textbf{d}_1,\textbf{d}_2,\textbf{d}_3>) \in \|r_B \|$, so that
the function $f = \alpha(q_{A,i})$ represents the transferring of
the tuples in relations of the source instance databases into the
target instance database $B = \alpha^*(\B)$, according to the SOtgd
$\Phi$ of the mapping $\M_{AB}
= \{\Phi\}:\A \rightarrow \B$.\\
In this way, for a given R-algebra $\alpha$ which satisfies the
conditions for the mapping-interpretations in Definition
\ref{def:mapping-interpretation}, we translate a logical
representation of database mappings, based on SOtgds, into an
algebraic representation based on relations of the instance
databases and the functions obtained from mapping-operads.
\\$\square$\\
It is easy to verify that for a \emph{query mapping}
$~\phi_{Ai}(\textbf{x}) \Rightarrow r_B(\textbf{t})$, a
mapping-interpretation $\alpha$ is an R-algebra such that the
relation $\alpha(r_q)$ is just equal to the image of the function
$\alpha(q_{A,i})$. The mapping-interpretation of $v_i$ is the
transfer of information of this computed query into the relation
$\alpha(r_B)$ of the database $\B$. \\When $\alpha$ satisfies this
query mapping $\phi_{Ai}(\textbf{x}) \Rightarrow r_B(\textbf{t})$,
then  $\alpha(r_q)  \subseteq \alpha(r_B)$ and, consequently, the
function $\alpha(v_i)$ is an injection, i.e., the \emph{inclusion}
of $\alpha(r_q)$ into $\alpha(r_B)$.\\
Moreover, each R-algebra $\alpha$ of a given set of mapping-operads
between a source schema $\A$ and target schema $\B$ determines a
particular information flux from the source
 into the target schema.
\begin{definition} \textsc{Information Flux} \label{prop:influx}\\
Let  $\alpha$ be a mapping-interpretation
\index{mapping-interpretations} (an R-algebra in Definition
\ref{def:mapping-interpretation}) of a given set
$\textbf{\emph{M}}_{AB} = \{q_1,...,q_n, 1_{r_\emptyset} \} =
MakeOperads(\M_{AB})$ of mapping-operads, obtained from an
\textsl{atomic} mapping $\M_{AB}:\A \rightarrow \B$, and $A =
\alpha^*(S_A)$ be an instance of the schema $\A = (S_A, \Sigma_A)$
that satisfies all constraints in $\Sigma_A$.
\\For each
operation $q_i \in \textbf{\emph{M}}_{AB}$, $~q_i = (e \Rightarrow
(\_~)(\textbf{t}_i)) \in O(r_{i,1},...,r_{i,k}, r'_i)$, let
$\textbf{x}_i$ be its tuple of variables which appear at least one
time free (not as an argument of a function) in $\textbf{t}_i$ and
appear as variables in the
atoms of relational symbol of the schema $\A$ in the formula $e[(\_~)_j/r_{i,j}]_{1\leq j \leq k}$. Then, we define\\
(1) $~~Var(\textbf{\emph{M}}_{AB}) =
\bigcup_{1\leq i \leq n} \{\{x\} ~|~ x \in \textbf{x}_i \}$.\\
We define the kernel of the information flux \index{information
flux} of
$\textbf{\emph{M}}_{AB}$, for a given mapping-interpretation $\alpha$, by (we denote the image of a function $f$ by '$im(f)$')\\
(2) $~~\Delta(\alpha, \textbf{\emph{M}}_{AB}) = \{\pi_{\textbf{x}_i}
(im(\alpha(q_i)))
 ~|~ q_i   \in \textbf{\emph{M}}_{AB}$,
  and $~\textbf{x}_i$ is not empty $\} \bigcup \perp^0$,
 $~~~~$ if $~Var(\textbf{\emph{M}}_{AB})\neq \emptyset$; $~\perp^0~$ otherwise.\\
We define the information flux by its kernel by\\
(3)$~~Flux(\alpha, \textbf{\emph{M}}_{AB}) = T(\Delta(\alpha,
\textbf{\emph{M}}_{AB}))$. \\
The flux of composition of $~\textbf{\emph{M}}_{AB}$ and
$\textbf{\emph{M}}_{BC}$ is
defined by:\\
$Flux(\alpha, \textbf{\emph{M}}_{BC} \circ \textbf{\emph{M}}_{AB}) =
Flux(\alpha, \textbf{\emph{M}}_{AB}) \bigcap Flux(\alpha,
\textbf{\emph{M}}_{BC})$.\\
(4)$~~$ We say that an information flux is \textsl{empty} $~$ if it
is equal to $\perp^0 = \{\perp\}$ (and hence it is not the empty
set), analogously as for an empty instance-database.
\end{definition}
 The information flux of the
SOtgd of the mapping $\M_{AB}$  for the instance-level mapping  $f =
\alpha^*(\textbf{M}_{AB}):A\rightarrow \alpha^*(\B)$ composed of the
set of functions $f = \alpha^*(\textbf{M}_{AB}) =
\{\alpha(q_1),...,\alpha(q_n), q_\perp \}$, is denoted by
$\widetilde{f}$. Notice that  $\perp \in \widetilde{f}$, and hence
the information flux $\widetilde{f}$
is a instance-database as well.\\
 From this definition,
each instance-mapping is a set of functions whose information flux
 is the intersection of the information
fluxes of all atomic instance-mappings that compose this composed
instance-mapping. These basic properties of the instance-mappings is
used in order to define the database $\textbf{DB}$ category where
the instance-mappings are the morphisms (i.e., the arrows) of this
category, while the instance-databases (each instance-database is a
set of relations of a schema also with the empty relation $\perp$)
are its objects.\\
 \textbf{Equality of morphisms}: The fundamental property
in $\textbf{DB}$ is the
following:\\
Any two arrows $f,g:A \rightarrow B$  where $A$ and $B$ are the
instance databases (the simple sets of the relations) in
$\textbf{DB}$ are \emph{equal} if $\widetilde{f} = \widetilde{g}$,
i.e., the have the same information fluxes.
\section{Saturation of the morphisms in DB}
Let $\phi_{Ai}(\textbf{x}) \Rightarrow r_B(\textbf{t})$, as in
Definition \ref{def:mapping-interpretation}, be an implication
$\chi$ in a normalized SOtgd $\exists \textbf{f}( \Psi)$ (where
$\Psi$ is a FOL formula) of the mapping $\M_{AB}:\A \rightarrow B$
with the sketch's arrow $\textbf{M}_{AB}= MakeOperads(\M_{AB}) =
\{q_1,...,q_n, 1_{r_\emptyset}\}$, $\textbf{t} = \langle
t_1,...,t_{ar(r_B)}\rangle$ be a tuple of terms with variables in
$\textbf{x} = <x_1,...,x_m>$, and  $q_i \in \textbf{M}_{AB}$ be the
operad's operation of this implication, equal to the expression $(e
\Rightarrow (\_~)(\textbf{t}))\in O(r_1,...,r_k,r_B)$, where $q_i =
v_i \cdot q_{A,i}$ with $q_{A,i} = (e \Rightarrow (\_~)(\textbf{t}))
\in O(r_1,...,r_k,r_q)$ and $v_i = ((\_~)(y_1,...,y_{ar(r_B)})\\
\Rightarrow (\_~)(y_1,...,y_{ar(r_B)})) \in O(r_q, r_B)$ such that
 for a new relational symbol $r_q$,$~ar(r_q) = ar(r_B) \geq 1$.\\
It is important to underline that each term $t_i$ is a simple
variable  which appear in the tuple $\textbf{x}$ (left side of the
implication) or the term $f_{l}(\textbf{z})$ where the variables in
the tuple $\textbf{z}$ is a subset of the variables in $\textbf{x}$.\\
For a given mapping-interpretation $\alpha$ such that $A =
\alpha^*(\A)$ and $B = \alpha^*(\B)$ are two models of the schemas
$\A$ and $\B$ respectively, and $\alpha$ satisfies the schema
mapping $\M_{AB} = \{\exists \textbf{f}\Psi \}$, with  the tuple of
existentially quantified Skolem functions $\textbf{f}$, the process
of saturation of the morphism $h = \alpha^*(\textbf{M}_{AB}) = \{
\alpha(q_1),...,\alpha(q_n), q_{\perp} \}$ is relevant only for the
operads operations $q_i$ which have at least one functional symbol
of $\textbf{f}$ in on the right side of implication, as follows:
\\\\
\underline{\textbf{Saturation algorithm} $Sat(\alpha^*(\textbf{M}_{AB}))$}\\\\
\textbf{Input}: A  mapping arrow $\textbf{M}_{AB} = \{q_1,...,q_N,
1_{r_\emptyset} \}:\A \rightarrow \B$, and a mapping-interpretation
$\alpha$ such that $A = \alpha^*(\A)$ and $B = \alpha^*(\B)$ are two
models of the schemas $\A$ and $\B$ respectively, and $\alpha$
obtained of a given Tarski's interpretation $I_T$, satisfies the
schema mapping $\M_{AB} = \{\exists \textbf{f}\Psi
\}$, with  the tuple of existentially quantified Skolem functions $\textbf{f}$.\\
\textbf{Output}: Saturated morphism   from $A$ into $B$ in
\textbf{DB} category.
\begin{enumerate}
\item Let $h = \alpha^*(\textbf{M}_{AB})$. Then initialize  $Sat(h) = h$, i = 0.
\item $i = i +1$. If $i > N$ go to 8.
\item Let the mapping component $q_i \in \textbf{M}_{AB}$ be the
 expression $(e \Rightarrow (\_~)(\textbf{t}))\in
O(r_1,...,\\r_k,r_B)$, where $q_i = v_i \cdot q_{A,i}$ with $q_{A,i}
= (e \Rightarrow (\_~)(\textbf{t})) \in O(r_1,...,r_k,r_q)$ and $v_i
= ((\_~)(y_1,...,y_{ar(r_B)}) \Rightarrow
(\_~)(y_1,...,y_{ar(r_B)})) \in O(r_q, r_B)$. Define the set $\F
\subseteq \textbf{f}$ of all functional symbols in the tuple of
terms $\textbf{t}$. If $\F$ is empty then go to 2.
\item (\emph{Fix the function of} $q_i$ for given $\alpha$)
 Let $f= \alpha(q_{A,i}):R_1 \times ...\times R_k \rightarrow
\alpha(r_q)$ be the function of this mapping-interpretation provided
in Definition \ref{def:mapping-interpretation} where $\alpha(r_q)
\subseteq \|r_B\|$ is image of this function with relation $\|r_B\|
= \alpha(r_B) \in B$,  $\textbf{x} = <x_1,...,x_m>$ be the tuple of
all variables in the left-side expression $e$ of the operad's
operation $q_i$, and $S$ be the set of sets that contain the pairs
of mutually equal free variables in the formula
$e[(\_~)_n/r_n]_{1\leq n \leq k}$ obtained from $q_i$ (in Definition
\ref{def:mapping-interpretation}).\\
Set $R_L = R_1 \times ...\times
 R_k$.
\item (\emph{Expansion of} $q_{A,i}$) If  $R_L$ is empty then go to 2.\\
 Take a tuple $\langle \textbf{d}_1,...,\textbf{d}_k\rangle\in
R_L \subseteq R_1 \times ...\times
 R_k$ and delete it from $R_L$. Then define the assignment $g:\{x_1,...,x_m\} \rightarrow \D$ such that
  $\langle g(x_1),...,g(x_m)\rangle = Cmp(S,\langle
 \textbf{d}_1,...,\textbf{d}_k\rangle$ (from Definition
 \ref{def:mapping-interpretation}).\\
 If $f(\langle \textbf{d}_1,...,\textbf{d}_k\rangle) = g^*(\textbf{t}) = \langle
 g(t_1),...,g(t_{ar(r_B)}\rangle \neq <>$ then go to 6.\\
Go to 5.
\item (\emph{Definition of the extension corresponding  to the tuple} $\langle
\textbf{d}_1,...,\textbf{d}_k\rangle$)\\
Let $Z$ be the set of indexes of the terms in $\textbf{t} = \langle
t_1,...,t_{ar(r_B)} \rangle$ which are simple variables and we
denote by $nr_{r_B}(j)$ the name of the j-th column of the relation
$r_B \in \B$. Then we define the relation:\\
 $R = ($SELECT $(*)$ FROM $\|r_B\|$ WHERE
$\bigwedge_{j \in Z}(nr_{R_B}(j) = g(t_j))) \backslash
\{g^*(\textbf{t}) \}$.
\item If $R$ is empty relation then go to 5.\\
Take from $R$ a tuple $\textbf{b} = \langle b_1,...,b_{ar(r_B)}
\rangle$ and delete it from $R$.
 We define a new Tarski's
 interpretation $I'_T$, different from $I_T$ only for
 the functional symbols $f_l \in \F$ of the j-th term $t_j = f_l(x_{j1},...,x_{jp}) \in
 \textbf{t}$, as follows:\\
 1. $I'_T(f_l)(g(x_{j1}),...,g(x_{jp}))  = b_j \neq
 g(t_j) = I_T(f_l)(g(x_{j1}),...,g(x_{jp}))$;\\
 2. For all assignments $g_1 \neq g$ we have that\\
  $I'_T(f_l)(g_1(x_{j1}),...,g_1(x_{jp})) =
  I_T(f_l)(g_1(x_{j1}),...,g_1(x_{jp}))$;\\
  so that for the R-algebra $\alpha'$ derived from the Tarski's
  interpretation $I'_T$, we obtain the new function $f_{\textbf{b}} =
  \alpha'(q_{A,i})$ which satisfies $f_{\textbf{b}}(\langle
  \textbf{d}_1,...,\textbf{d}_k\rangle) = \textbf{b}$.\\
     Insert the function $f_{\textbf{b}}:R_1 \times ...\times R_k \rightarrow
\|r_B\|$ in $Sat(h)$ and go to 7.
\item   \textbf{Return} the saturated morphism  $Sat(\alpha^*(\textbf{M}_{AB})):A \rightarrow B$.
   \end{enumerate}
Notice that for a  mapping sketch's arrow $\textbf{M}_{AB} =
\{q_1,...,q_N, 1_{r_\emptyset} \}:\A \rightarrow \B$, and a
mapping-interpretation $\alpha$ (such that $A = \alpha^*(\A)$ and $B
= \alpha^*(\B)$ are two models of the schemas $\A$ and $\B$,
respectively, and $\alpha$, obtained of a given Tarski's
interpretation $I_T$, satisfies the schema mapping $\M_{AB}$), we
obtain the $\textbf{DB}$ morphism $h = \alpha^*(\textbf{M}_{AB}) =
\{\alpha(q_1),...,\alpha(q_N),q_\bot\}:A \rightarrow B$ with the
property that each k-ary function $\alpha(q_i):R_1\times...\times
R_k \rightarrow \|r_B\|$ for its argument returns a single tuple (or
empty tuple
$<>$) of $\|r_B\|$.\\
Let $dom$ and $cod$ be the operators which, for each function,
return the domain and codomain of this function, respectively, and
$\P$ be the powerset operation.  By the saturation of $h$ we obtain
the morphism $Sat(h):A \rightarrow B$ from which we are able to
define the  set $S_{q_i} = \{h_i \in Sat(h) ~|~ dom(h_i) =
dom(\alpha(q_i))$ and $cod(h_i) = cod(\alpha(q_i)) \}$ and function
$f_{q_i} =  \bigcup S_{q_i}  \triangleq \bigcup_{h_i \in S_{q_i} }
graph(h_i)$, where\\
 ($\mho$) $~~~~graph(h_i) = \{(\langle
  \textbf{d}_1,...,\textbf{d}_k\rangle), h_i(\langle
  \textbf{d}_1,...,\textbf{d}_k\rangle)~|~\langle
  \textbf{d}_1,...,\textbf{d}_k\rangle \in dom(h_i)$ and $h_i(\langle
  \textbf{d}_1,...,\textbf{d}_k\rangle \neq <> \}$\\ is the non-empty-tuple graph
  of this function. Thus, in this way we obtain the p-function:\\
  ($\wp$) $~~~~f_{q_i}:dom(\alpha(q_i)) \rightarrow \P(cod(\alpha(q_i))$,\\
   for  each operad's operation $q_i \in \textbf{M}_{AB}$ which has the
  functional symbols on he right side of implication in $q_i$.\\
  We have the following property for these derived p-functions:
  \begin{lemma} \label{lemma:powerset}
  Let the mapping component $q_i \in \textbf{\emph{M}}_{AB}:\A \rightarrow \B$ be the
 expression $(e \Rightarrow (\_~)(\textbf{t}))\in
O(r_1,...,r_k,r_B)$, with  the tuple $\textbf{x} = <x_1,...,x_m>$
of all variables in the left-side expression $e$,  such that the set
 of  functional symbols in the tuple of terms $\textbf{t}$ is not empty.
Let  R-algebra $\alpha$ be a model of this mapping
$\textbf{\emph{M}}_{AB}$ with $R_i = \alpha(r_i) \in A =
\alpha^*(\A)$, $i = 1,...,k$, and $\|r_B\| = \alpha(r_B) \in B =
\alpha^*(\B)$, and $ (\alpha(q_i):R_1\times...\times R_k \rightarrow
\|r_B\|) \in h = \alpha^*(\textbf{\emph{M}}_{AB}):A \rightarrow
B$.\\ Then, for the set $S_{q_i} = \{h_i \in Sat(h) ~|~ dom(h_i) =
dom(\alpha(q_i))$ and
$cod(h_i) = cod(\alpha(q_i)) \}$ we define the p-function:\\
 (1) $~~~f_{q_i} = \bigcup S_{q_i}  \triangleq \bigcup_{h_i \in S_{q_i} }
graph(h_i):R_1\times...\times R_k \rightarrow \P(\|r_B\|)$.\\
Let $Z$ be the set of indexes of the terms in $\textbf{t} = \langle
t_1,...,t_{ar(r_B)} \rangle$ which are simple variables and we
denote by $~nr_{r_B}(j)$ the name of the j-th column of the relation
$r_B \in \B$. Then, for each tuple $\langle
\textbf{d}_1,...,\textbf{d}_k\rangle\in  R_1 \times ...\times
 R_k$ with the assignment $g:\{x_1,...,x_m\} \rightarrow \D$ such that
  $\langle g(x_1),...,g(x_m)\rangle = Cmp(S,\langle
 \textbf{d}_1,...,\textbf{d}_k\rangle$ (from Definition
 \ref{def:mapping-interpretation}), we obtain:\\
 (2) $~f_{q_i}(\langle  \textbf{d}_1,...,\textbf{d}_k\rangle) =
 ~$\textsc{SELECT} $(*)$ \textsc{FROM} $\|r_B\|$ \textsc{WHERE}
$\bigwedge_{j \in Z}(nr_{R_B}(j) = g(t_j))$,\\
and, if $~\alpha(q_i)(\langle  \textbf{d}_1,...,\textbf{d}_k\rangle)
= <>~$ then $~f_{q_i}(\langle  \textbf{d}_1,...,\textbf{d}_k\rangle)
= \emptyset \in \P(\|r_B\|)$.
  \end{lemma}
\textbf{Proof}: From the step 6 and 7 of the algorithm for
saturation, we have that for every tuple in $R = $\textsc{SELECT}
$(*)$ \textsc{FROM} $\|r_B\|$ \textsc{WHERE} $\bigwedge_{j \in
Z}(nr_{R_B}(j) = g(t_j))$, we have one function in the set
$S_{q_i}$, and consequently the equation (2) is valid.
\\$\square$
\begin{coro} \label{coro:saturation}
For every R-algebra $\alpha$ which is a model of a given schema
mapping $\textbf{\emph{M}}_{AB}:\A \rightarrow \B$, we have that
$\alpha^*(\textbf{\emph{M}}_{AB}):\alpha^*(\A) \rightarrow
\alpha^*(\B)$ and
$Sat(\alpha^*(\textbf{\emph{M}}_{AB})):\alpha^*(\A) \rightarrow
\alpha^*(\B)$ are two \verb"equal" morphisms in the category
$\textbf{DB}$.\\
Consequently, the saturation of morphisms is an invariant process in
$\textbf{DB}$, so that we can replace any non-saturated morphisms
with its saturated version in any commutative diagram in
$\textbf{DB}$.
\end{coro}
\textbf{Proof}: From the fact that the introduction of the new
functions changes only the terms with non-built-in functional
symbols on the right sides of implications, so that they are not in
$Var(\textbf{M}_{AB})$, and hence, from Definition
\ref{prop:influx}, they do not change the information flux of the
morphism $\alpha^*(\textbf{M}_{AB})$.
\\$\square$\\
\textbf{Example 4}: Let us consider the following simple example
with three relations in the database schema $\A$:\\
\begin{figure}
 \includegraphics[scale=0.98]{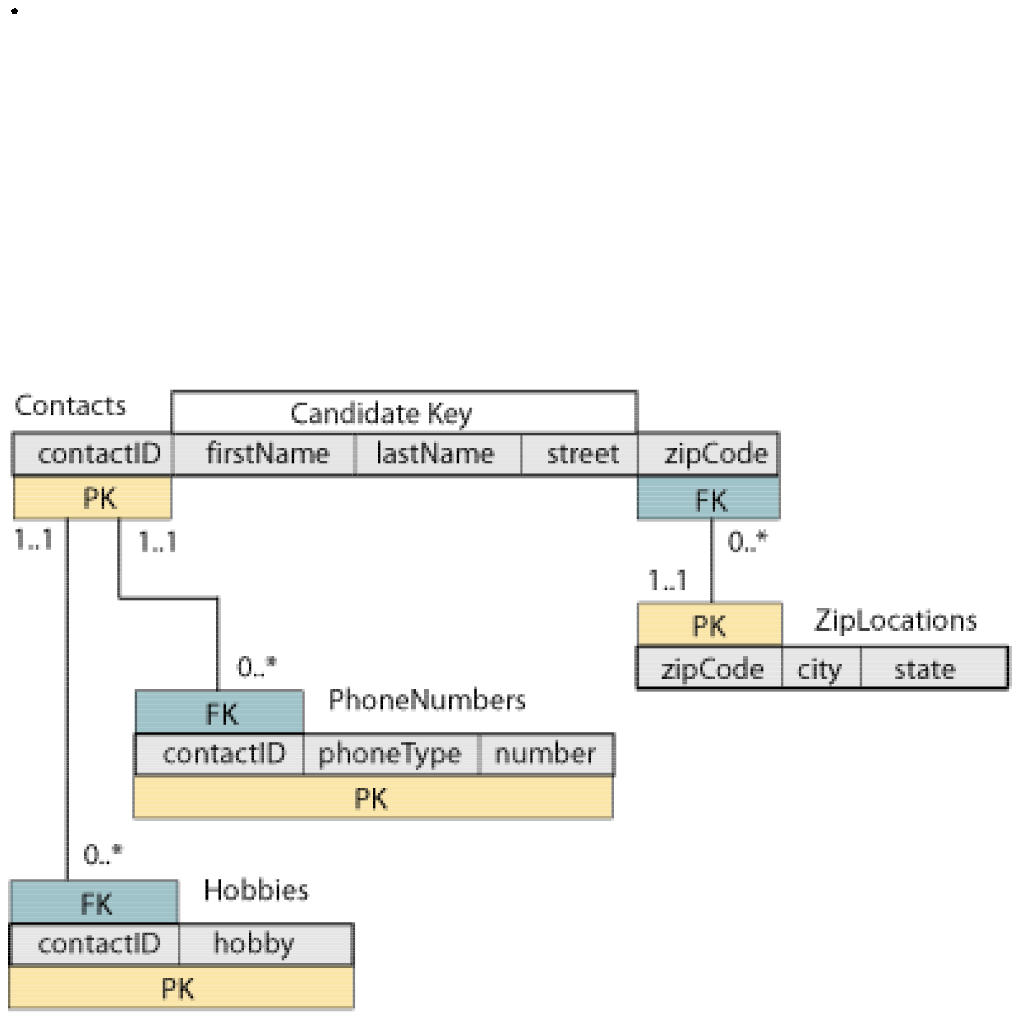}
   \label{fig:treeinf}
  $\vspace*{-28mm}$
 \end{figure}
1. $\verb"ZipLocations"(\verb"zipCode",\verb"city",\verb"state")$
with
primary key (PK)\verb"zipCode",\\
2.
$\verb"Contacts"(\verb"contactID",\verb"firstName",\verb"lastName",\verb"street",
\verb"zipCode")$ with PK corresponding to \verb"contactID" and
foreign key (FK) to \verb"zipCode", and\\
3.$\verb"PhoneNumbers"(\verb"contactID",\verb"phoneType",\verb"number")$
with FK \verb"contactID",\\ such that for each contact we can store
the name and forename of the contacted person, his address and phone
numbers. Suppose that we want to know what hobbies each person on
our contact list is interested in. It can be only done
\emph{indirectly} by introducing a database schema $\B$ with a
relation $\verb"Hobbies"(\verb"contactID",\verb"hobby")$ with FK
\verb"contactID", and hence represented by  the schema above.\\
%
Consequently, we define a schema mapping $\M_{AB}:\A \rightarrow \B$
by the tgd $\forall
x_1,x_2,x_3,x_4,x_5\\(\verb"Contacts"(x_1,x_2,x_3,x_4,x_5)
\Rightarrow \exists y \verb"Hobbies"(x_1,y))$, so that by
Skolemization we obtain the SOtgd  $\Phi$ equal to the logic
formula\\ $\exists f_1(\forall
\textbf{x}(\verb"Contacts"(x_1,x_2,x_3,x_4,x_5) \Rightarrow
\verb"Hobbies"(x_1,f_1(x_1)))$, where \\$\textbf{x} = \langle
x_1,x_2,x_3,x_4,x_5 \rangle$. Consequently, $\textbf{M}_{AB} =
MakeOperads(\{\Phi\}) = \{q_1, 1_{r_\emptyset}\}:\A \rightarrow \B$,
with $q_1 = v_1 \cdot q_{A,1}\in O(\verb"Contacts",\verb"Hobbies")$
with $q_{A,1} = ((\_)(\textbf{x}) \Rightarrow (\_)(\textbf{t})) \in
O(\verb"Contacts",r_q)$, where $\textbf{t} = \langle t_1,t_2\rangle$
with the term $t_1$ equal to variable $x_1$ and term $t_2$ equal to
$f_1(x_1)$, and $v_1 = ((\_)(y_1,y_2) \Rightarrow (\_)(y_1,y_2)) \in
O(r_q,\verb"Hobbies")$.\\
Let us consider a model of this schema mapping $\alpha$, such that:
$R_1 = \alpha(\verb"Contacts")$ and $\|r_B\| = \|\verb"Hobbies"\| =
\alpha(\verb"Hobbies")$, with
 \\\\
$R_1 =$
\begin{tabular}{|lllll|}
  \hline
  \verb"contactID" & ~\verb"firstName" & ~\verb"lastName" & ~\verb"street" & ~\verb"zipCode"   \\
  \hline
   ... & ~... & ~... & ~... & ~...  \\
  132 & ~Zoran & ~Majkic & ~Appia & ~0187  \\
  ... & ~... & ~... & ~... & ~...  \\
    \hline
\end{tabular}\\\\\\
$\|r_B\| = \|\verb"Hobbies"\| =$ \begin{tabular}{|ll|}
  \hline
    ~\verb"contactID" & ~\verb"hobby" \\
  \hline
  ~... & ~...\\
   ~132 & ~photography\\
  ~132 & ~music\\
   ~132 & ~art\\
   ~132 & ~travel\\
   ~... & ~...\\
  \hline
\end{tabular} \\\\\\
so that for $\textbf{d}_1 = \langle 132,Zoran,Majkic,Appia,0187
\rangle \in R_1$, we obtain the assignment
$g:\{x_1,x_2,x_3,x_4,x_5\} \rightarrow \D$ such that $\langle
g(x_1),g(x_2),g(x_3),g(x_4),g(x_5)\rangle =
Cmp(\emptyset,\textbf{d}_1) = \textbf{d}_1$, i.e, $g(x_1) = 132,
g(x_2) = Zoran, g(x_3) = Majkic, g(x_4) = Appia$ and $g(x_5) =
0187$, and for $f = \alpha(q_{A,1}):R_1 \rightarrow \alpha(r_q)$
such that $f(\textbf{d}_1) = g^*(\textbf{t}) = \langle
g(x_1),g(f_1(x_1))\rangle = \langle 132, I_T(f_1)(132)\rangle =
\langle 132, art \rangle$, that is $I_T(f_1)(132) = art$.\\
Then in step 6 of the algorithm, we have that $Z = \{t_1\} = \{x_1\}$
with $nr_{\verb"Hobbies"}(1) = \verb"contactID"$ and\\
$R = ($SELECT $(*)$ FROM $\|r_B\|$ WHERE $\bigwedge_{j \in
Z}(nr_{R_B}(j) = g(t_j))) \backslash \{g^*(\textbf{t}) \}\\
= ($SELECT $(*)$ FROM $\|r_B\|$ WHERE $contactID = 132)) \backslash
\{g^*(\textbf{t}) \}$
\\\\
$ =$ \begin{tabular}{|ll|}
  \hline
    ~\verb"contactID" & ~\verb"hobby" \\
  \hline
     ~132 & ~photography\\
  ~132 & ~music\\
     ~132 & ~travel\\
   \hline
\end{tabular} \\\\
Consequently, in step 7 of the algorithm will be introduced the
three new functions from $R_1$ into $\|r_B\|$, $f_{\textbf{b}},
\textbf{b} \in R$, into $Sat(\alpha^*(\textbf{M}_{AB}))$ so that\\
$f_{132,photograpy}(\textbf{d}_1)  = \langle 132,photography
\rangle$, with $I'_T(f_1)(132) = photography$;\\
$f_{132,music}(\textbf{d}_1)  = \langle 132,music \rangle$, with $I'_T(f_1)(132) = music$;\\
$f_{132,travel}(\textbf{d}_1)  = \langle 132,travel \rangle$, with $I'_T(f_1)(132) = travel$.\\
Thus, for the derived function $f_{q_1} = \bigcup
S_{q_1}:\alpha(\verb"Contacts")
\rightarrow\P(\alpha(\verb"Hobbies"))$, we obtain that\\
$f_{q_1}(\textbf{d}_1) = f_{q_1}(132,Zoran,Majkic,Appia, 0187) =
\\\\
 =$ \begin{tabular}{|ll|}
  \hline
    ~\verb"contactID" & ~\verb"hobby" \\
  \hline
     ~132 & ~art\\
     ~132 & ~photography\\
  ~132 & ~music\\
     ~132 & ~travel\\
   \hline
\end{tabular} \\\\
and hence, by using the second projection $\pi_2$, we obtain that\\
$(\pi_2 \cdot f_{q_1})(\textbf{d}_1) = \{photography, art, music,
travel\}$,\\
that is, for each contact ID, the function $\pi_2 \cdot f_{q_1}$
returns the set of hobbies of this ID.
\\$\square$\\
In this way we are able to represent also the 1:N relationships
between relational tables by the morphisms in $\textbf{DB}$
category.\\
It is important that the saturation can be done only for the non
built-in functional symbols. In fact we have only one prefixed
interpretation of the built-in functional symbols, so that their
interpretation is equal for every Tarski's interpretation. Let us
show one  example with functional symbols that are built-in
functions:\\\\
\textbf{Example 5}: Let us consider the IRDB with the parsing of the
RDB instances into the vector relation
$r_V(\verb"r-name",\verb"t-index",\verb"a-name",\verb"value")$,
introduced in \cite{Majk14R,Majk14I},where is demonstrated the
following proposition:
\begin{itemize}
  \item Let the IRDB be given by a Data Integration system $\I = \langle
\A,\S,\M \rangle$ for a used-defined global  schema $\A =
(S_A,\Sigma_A)$ with $S_A =\{r_1,...,r_n\}$, the source schema $\S =
(\{r_V\},\emptyset)$ with the vector big data relation $r_V$ and the
set of mapping tgds $\M$ from the source schema into  he relations
of the global schema. Then  a canonical model of $\I$ is any model
of the schema $\A^+ = (S_A \bigcup \{r_V\}, \Sigma_A \bigcup \M
\bigcup \M^{OP})$, where $\M^{OP}$ is an opposite mapping tgds  from
$\A$ into $r_V$  given by the following set of tgds:\\
$\M^{OP} = \{\forall x_1,...,x_{ar(r_k)}((r_k(x_1,...,x_{ar(r_k)})
\wedge x_i \textsc{NOT NULL}) \Rightarrow\\
r_V(r_k, Hash(x_1,...,x_{ar(r_k)}),nr_{r_k}(i),x_i))~|~1\leq i \leq
ar(r_k), r_k \in S_A \}:\A \rightarrow \S$.
\end{itemize}
 Thus, $\textbf{M}^{OP} =
MakeOperads(\M^{OP}) = \{1_{r_\emptyset}\} \bigcup \{q_{k,i}| r_k
\in S_A$ and $1 \leq i \leq ar(r_k)\}:\A \rightarrow \S$ is a
sketch's mapping with $q_{k,i} =
(((\_)(x_{k,1},...,x_{k,ar(r_k)})\wedge x_{k,i} \textsc{NOT NULL})
\Rightarrow (\_)(\textbf{t}_{k,i})) \in O(r_k,r_V)$, where
$\textbf{t}_{k,i}
= \{t_1,...,t_4\}$ with the terms:\\
1. $t_1$ is the nullary built-in function, i.e., the fixed constant
which does not depend on Tarski's interpretations, equal to the
relation table name $r_k$;\\
2. $t_2 = Hash(x_{k,1},...,x_{k,ar(r_k)})$ where $Hash$ is a built
in-function equal for every Tarski's interpretation;\\
3. $t_3$ is the nullary built-in function, i.e., the fixed constant
which does not depend on Tarski's interpretations, equal to the i-th
column name of the relational table $r_k$;\\
4. $t_4$ is the variable $x_{k,i}$.\\
Thus, no one of these three built-in functional symbols are obtained
by elimination of the existentially quantified variables, so they
are not the Skolem functions, and in the SOtgd of $\M^{OP}$ the set
of existentially quantified functional symbols $\textbf{f}$ is
empty, so that from the algorithm of saturation, for a given
R-algebra $\alpha$, such that $A = \alpha^*(\A)$ is the instance
database of the schema $\A$ and $\overrightarrow{A} = \alpha(r_V)$
is the obtained vector relation by parsing the database $A$, we
obtain that $Sat(\alpha^*(\textbf{\emph{M}}^{OP})) =
\alpha^*(\textbf{\emph{M}}^{OP})$.\\
We recall that the operation of parsing, \underline{PARSE}, for a
tuple $\textbf{d} =\langle d_1,...,d_{ar(r_k)}\rangle$ of the
relation $R_k = \|r_k\| = \alpha(r_k) \in A$,
is defined by the mapping\\
 $(r_k,\textbf{d}) ~~\mapsto ~~\{\langle r_k,
 Hash(\textbf{d}),nr_{r_k}(i),d_i \rangle |~  d_i \textsc{NOT NULL},
1\leq i \leq ar(r_k)\}$,
so that\\
 $~~~\overrightarrow{A} = \bigcup_{r_k \in S_A,\textbf{d} \in
 \|r_k\|} \textsc{\underline{PARSE}}(r_k,\textbf{d})$.\\
 Consequently, we obtain the function $\alpha(q_{k,i}):\alpha(r_k)
\rightarrow \alpha(r_V) = \overrightarrow{A}$, such that for its
image $im(\alpha(q_{k,i}))$ we obtain that from the parsing
$\pi_{4}(im(\alpha(q_{k,i}))) = \pi_i(\alpha(r_k))$.\\
If we make union of all functions in $f^{OP} =
\alpha^*(\textbf{\emph{M}}^{OP})$ with the same domain and codomain,
for example, for the domain $R_k = \alpha(r_k) \in A$, we obtain the
p-function\\
$f_{r_k} = \bigcup_{1\leq i \leq ar(r_k} \alpha(q_{k,i}):R_k
\rightarrow \P(\overrightarrow{A})$,\\
such that for each tuple $\textbf{d} =\langle
d_1,...,d_{ar(r_k)}\rangle$ of the relation $R_k = \|r_k\| =
\alpha(r_k) \in A$, \\
$f_{r_k}(\textbf{d}) =
\textsc{\underline{PARSE}}(r_k,\textbf{d})$.\\
If we use the schema $\A$ in Example 4, and $r_k = \verb"Contacts"$,
then for the tuple $\textbf{d} = \langle
132,Zoran,Majkic,Appia,00187 \rangle$, we obtain:\\\\
$f_{r_k}(\textbf{d}) = \textsc{\underline{PARSE}}(r_k,\textbf{d}) =
 $ \begin{tabular}{|llll|}
  \hline
    ~\verb"r-name" & ~\verb"t-index" & ~\verb"a-name"& ~\verb"value"\\
  \hline
     ~\verb"Contacts" & ~IND &~\verb"contactID" &~132\\
     ~\verb"Contacts" & ~IND &~\verb"firstName" &~Zoran\\
  ~\verb"Contacts" & ~IND &~\verb"lastName" &~Majkic\\
     ~\verb"Contacts" & ~IND &~\verb"street" &~Appia\\
     ~\verb"Contacts" & ~IND &~\verb"zipCode" &~0187\\
   \hline
\end{tabular} \\\\
where $\textsc{IND} = Hash(132,Zoran,Majkic,Appia,00187)$.\\
 Consequently, the parsing can be derived from the morphism in
$\textbf{DB}$ category,\\ $f^{OP}:A \rightarrow
\{\overrightarrow{A}, \bot\} = \alpha^*(\S)$.
\\$\square$
\section{Conclusion}

It was demonstrated that a categorical logic (denotational
semantics) for database schema mapping based on views is a very
general framework for RDBs, the
 database-integration/
 exchange and peer-to-peer systems \cite{Majk14}.
 In this very general semantic framework  was necessary to introduce the
 base database category $~\textbf{DB}~$  (instead of traditional $\textbf{Set}$ category),
 with objects instance-databases and with
 morphisms (mappings which are not simple functions) between them,  at an \emph{instance level} as
  a proper semantic domain for a database mappings based on a set of complex query computations.\\
   The higher logical \emph{schema level} of mappings between databases, usually written in some
 high expressive logical language (ex.~\cite{Lenz02,FKMP03}, GLAV (LAV and GAV), tuple
 generating dependency)  can then be  translated functorially into this base
 "computation" category. Hence, the denotational semantics of database mappings is given by
   morphisms of the Kleisli category $\textbf{DB}_T$, based on the
 fundamental (from Universal algebra) monad (power-view endofunctor) $T$, which may be
   "internalized" in $\textbf{DB}$ category as "computations". Big Data integration
   framework presented in \cite{Majk14} considers  the standard
   RDBs with  Tarskian semantics of the FOL, where one defines what it takes for a
sentence in a language to be true relative to a model.\\
  In this paper we demonstrated that each morphisms in $\textbf{DB}$
  can be equivalently substituted by its saturation-morphism, and we have shown that in this way by the morphisms in
  $\textbf{DB}$ we are able to represent also the 1:N relationships
  between the relational tables, but also to define the parsing of
  the RDBs into intensional RDBs with the vector relations
  containing the data and the metadata (IRDBs). \\
  Moreover, the saturated morphisms are able to express the general
  mappings from any given tuple of some relational view (obtained by a given SQL
  statement) into the set of tuples of another relational tables,
  which generally can be used in intensional RDBS where we are using
  the intensional FOL with the extensionalization function for the
  intensional concepts. In a future work we will investigate these
  properties of saturated morphisms for more advanced features of the
  IRDBs as are the multivalued attributes (which can not be supported
  in the FOL and standard RDBs).


\bibliographystyle{IEEEbib}
\bibliography{mydb}



%
\end{document}